\documentclass[onecolumn, natbib]{mn2e}
\usepackage{epsf,graphics,graphicx}
\usepackage{amsmath}
\usepackage{amssymb,latexsym,mathrsfs, bm}
\usepackage{hyperref}
\usepackage{color}
\usepackage{float}

\def\bea{\begin{eqnarray}}
\def\eea{\end{eqnarray}}
\def\ba{\begin{array}}
\def\ea{\end{array}}

\def\beq{\begin{equation}}
\def\eeq{\end{equation}}

\title[Reconstruction of unlensed CMB spectra]{Towards reconstruction of unlensed, intrinsic CMB power spectra from lensed map}

\author[Barun Kumar Pal, Hamsa Padmanabhan and Supratik Pal]{Barun Kumar Pal$^{1}$\thanks{Electronic address: barunp1985@rediffmail.com},
Hamsa Padmanabhan$^2$\thanks{Electronic address: {hamsa@iucaa.ernet.in}},
  Supratik Pal$^1$\thanks{Electronic address: {supratik@isical.ac.in}
}\\
$^{1}$Physics and Applied Mathematics Unit, Indian Statistical Institute,
 Kolkata 700 108, India\\
$^{2}$ Inter-University Centre for Astronomy and Astrophysics, Pune 411007, India}

\begin{document}
\date{ }

\maketitle

\label{firstpage}

\begin{abstract}
We propose a method to extract the unlensed, intrinsic CMB temperature and polarization power spectra from the 
observed (i.e., lensed) spectra. Using a matrix inversion technique, we demonstrate how one can  reconstruct the intrinsic CMB power spectra 
directly from lensed data for both flat sky and full sky analyses. The delensed spectra obtained by the technique are calibrated against the Code for 
Anisotropies in the Microwave Background (CAMB) using WMAP 7-year best-fit data and applied to WMAP 9-year unbinned data as well. 
In principle, our methodology may help in subtracting out the $E$-mode lensing contribution in order to obtain the intrinsic $B$-mode power.
\end{abstract}


\section{Introduction}
Ever since the detection of temperature anisotropies in the Cosmic Microwave background (CMB) in 1992 by the COBE
satellite \cite{cobe1,cobe2}, the CMB has continued to 
surprise cosmologists and its prevalence is still outbreaking among different branches of physics. 
The heart of present-day observational 
cosmology dwells in the accurate measurement of CMB anisotropies. The survey of the CMB remains a very important 
tool to explore the physics of the early universe.
The latest observational probes like WMAP \cite{wmap7}, Planck \cite{planck},  ACT \cite{act}, SPT \cite{spt} 
have led to extremely precise data resulting in a construction of
a very accurate model of our universe. Even though the latest data from Planck show slight disagreement with the best-fit $\Lambda$CDM 
model for low multipoles ($\ell\lesssim 40$)
at $2.5\sigma - 3\sigma$, 
 the collective CMB data, so far as the entire regime of $\ell$ is concerned, are in excellent harmony with the
$\Lambda$CDM model and Gaussian adiabatic initial conditions with a slightly
red-tilted power spectrum for the primordial curvature perturbation. 
To facilitate the confinement of cosmological models as well as the cosmological parameters further,
we need more precise CMB measurements where secondary effects like weak gravitational lensing come into play.

All through the voyage from the last scattering surface (LSS) en-route to the present day detectors, the path of the 
CMB photons gets distorted by the potential 
gradients along the line of sight, this phenomenon is known as gravitational lensing. As a result of this lensing 
effect, the CMB temperature field is remapped. 
Due to gravitational lensing, the acoustic peaks in the CMB angular power spectrum are smoothed \cite{seljak1995} and 
the power
on smaller scales is enhanced \cite{linder1990,silk1997}. Accordingly, the lensing of the CMB becomes more and more important as on smaller scales where 
there is very little intrinsic power. The remapping of power of the CMB anisotropy spectra by the effect of lensing is a very small effect, but it is important 
in the current era of precision cosmology. In other words, subtraction of the lensing artifact is essential to obtain a precise handle on the physics at the last 
scattering surface.

In addition to the temperature anisotropies, the CMB is also linearly polarized, which was first detected by DASI \cite{dasi}. The observed polarization field is also 
lensed by the potential gradients along the line of sight. Consequently, the CMB radiation field  furnishes two additional lensed observables in the form of the $E$ 
and $B$ polarization modes. The primordial gravity waves generate (along with primordial magnetic
fields, if any) the large scale $B$-mode signal. The precise measurement of the primordial $B$-mode polarization, although yet to be observed (very recently 
SPT has claimed a detection of CMB $B$- mode polarization \cite{hanson} produced by gravitational lensing), is very crucial in the context of inflation as it is directly related 
to the inflationary energy scale \cite{knox1,knox2} (assuming there are no vector modes), which is essential to discriminate among different classes of inflationary models. 
However, there is a confusion between the CMB $E$ and $B$-modes in presence of lensing \cite{seljak}. The lensing of the CMB $E$-mode polarization 
produces a non-zero $B$-mode signal \cite{seljak1998}. Hence, the detection of a large scale $B$-mode signal in CMB polarization experiments does not ensure that we 
are actually observing primordial gravity waves, as it may be due to the lensing artifact. The lensing effect may also produce non-Gaussian features in the CMB maps 
\cite{bernardeau}. Hence, it is important to properly separate out the lensing effect from the observed CMB anisotropy spectra to recover the true features of 
primordial CMB radiation. (We note, however, that if a complete forward modelling approach is used to compare to the data, reconstruction of the lensing potential power spectrum alone is sufficient.) Given the form of the primordial curvature perturbation, one can work out the CMB angular power spectra as well as the
lensing potential power spectrum using codes such as CAMB.\footnote{A. Lewis and A. Challinor, URL: http://www.camb.info} Employing CAMB, the corresponding lensed CMB 
spectra can also be estimated. However, 
the available observational data \cite{planck,wmap7} directly provide the lensed CMB spectra, and hence, our aim is to reconstruct the unlensed quantities directly 
from the observed lensed ones.

In this article, we provide a simple algorithm for direct reconstruction of intrinsic CMB spectra from
the lensed ones by applying a matrix inversion technique.  Our technique assumes the availability of the lensing potential. 
Though we do not address the lensing potential reconstruction here, it may be done when complete polarization data (both $E$
and $B$ polarizations) are available.
We first construct the kernel matrix for the difference between the lensed and unlensed
CMB angular power spectra, which only depends on the knowledge of the power spectrum for the lensing potential. For our present analysis, 
we have used the WMAP 7-year best-fit spectra for $\Lambda$CDM + TENS model, since this is the latest available data for lensing as of now. 
Exploiting this matrix inversion technique, we are able to subtract out the lensing contribution from the 
CMB power spectra. As a preliminary test, we have also used the WMAP 9-yr unbinned $TT$ power spectrum and delensed by our matrix inversion technique, 
without taking into account the errors on the data.
The corresponding lensing potential is generated by CAMB using the WMAP 9-yr best-fit parameters for the $\Lambda$CDM + TENS 
model. 
Of course, this is just the first step towards extraction of 
unlensed CMB spectra, applicable under ideal conditions. 
For our theoretical 
framework to be applicable in the realistic situation, the uncertainties coming from the noise in the measured 
spectra and the transfer function must also be taken into account. 
This method, may,  in principle serve as an aid to resolve the confusion between the $E$ and $B$-mode once 
the CMB $B$-mode 
data is available and when a more realistic situation can be taken into account. The unlensed CMB spectra thus obtained, may 
also be helpful in increasing the level of precision in determining various cosmological parameters, 
though that potential consequence is not addressed in the present article. 

The paper is organized as follows. In Section \ref{reviw}, we provide a brief review of the theory of the weak 
gravitational lensing of the CMB,
which we will make use of in discussing the main body of the article in the subsequent sections.  In Section \ref{mit},
we provide our theoretical framework for the matrix inversion technique. The numerical results, and possible sources of error are discussed in Section \ref{num}. 
 We summarize our findings and discuss future prospects in a brief concluding section. 
\section{Lensing of the CMB}\label{reviw}

Weak gravitational lensing of the CMB photons occurs due to the intervening large-scale structure between the LSS and the observer. Due to the effect of weak lensing, a point $\hat{\bf n}$, on the LSS appears to be in a deflected position $\hat{\bf n}'$. 
The (lensed) temperature $\tilde{T}(\bf \hat{n})$ that we measure as coming from a direction $\bf \hat{n}$ in the sky, actually corresponds to the original 
(unlensed) temperature ${T}(\bf \hat{n}')$ from a different direction $\bf \hat{n}'$, where $\bf \hat{n}$ and $\bf \hat{n}'$ are related through the deflection angle $\bm{\alpha}$, as $\bf \hat{n}' = \bf \hat{n} + \bm{\alpha}$. 
Similarly, the polarization field is remapped according to $\tilde{P}(\hat{\bf n})=P(\hat{\bf n}')$.
The change in direction on the sky, $\bm{\alpha}$, can be related to the fluctuations of the gravitational potential, ${\Psi}$
\cite{seljak,durrer,CLreview06,challinor}.
In the linear theory, ${\Psi}$ can be related to the primordial curvature perturbation, $\mathcal{R}$, generated during inflation,
through the transfer function, $T(\eta; \mathbf{k})$, by the relation $\Psi(\eta,\mathbf{k})= T(\eta; \mathbf{k})\mathcal{R}(\mathbf{k})$. In terms of the primordial power spectrum $\mathcal{P}_{\mathcal{R}}(k)$, the angular
power spectrum of the lensing potential $\psi$, defined through $\bm{\nabla}\psi\equiv\bm{\alpha}$, is given by:
\beq
C_\ell^{\psi \psi}=16\pi\int\frac{dk}{k}\mathcal{P}_{\mathcal{R}}(k)\left[\int_0^{z_{rec}} \frac{dz}{H(z)}j_\ell(kz)
\frac{\chi(z_{rec})-\chi(z)}{\chi(z_{rec})\chi(z)}T(z, \mathbf{k})\right]^2
\label{pps}
\eeq
where $j_\ell(kz)$ is the spherical Bessel function of order $\ell$. Hence, given the form of the primordial power spectrum, the lensing potential power spectrum can be
 computed from Eq.\eqref{pps}. This can be done, for example, using numerical codes like CAMB.

\subsection{Lensed CMB Power spectra}
Given the lensing potential power spectrum, we can use it along with the unlensed temperature and polarization power spectra to derive the corresponding lensed quantities.
Here we briefly review the correlation function method to find the expressions for the lensed CMB
temperature and polarization power spectra, first assuming the flat-sky approximation and then in the full-sky limit. This correlation function technique was first
introduced in Ref. \cite{seljak1995} and later amended by others \cite{Lewis,CLreview06} with increased levels of 
accuracy. (An alternate approach, working entirely in harmonic space, is followed, e.g., in Ref. \cite{hu2000}.) In the present work, we shall follow the technique of Ref. \cite{Lewis}.

\vspace{0.cm}
\subparagraph*{\bf Flat-Sky Approximation:}
In the flat-sky approximation, the lensed temperature field is expressed in terms of a two-dimensional Fourier transform on the 
sky plane. It can be shown that the assumption of the deflection angle $\bm{\alpha}$ to be a Gaussian field leads to the following relation for the 
lensed temperature anisotropy correlation function \cite{seljak1995,Lewis}:
\beq
\tilde{\xi}(\theta)=\int\frac{d\ell'}{\ell'} \frac{\ell'^2C_{\ell'}^{TT}}{2\pi}e^{-\ell'^2\sigma^2(\theta)/2}\left[\left(1+\frac{1}{16}\ell'^4 A_2(\theta)^2\right)J_0( \ell' \theta)
+\frac{1}{2}\ell'^2 A_2(\theta)J_2(\ell'\theta)+\frac{1}{16}\ell'^4 A_2(\theta)^2 J_4(\ell'\theta )\right],
\label{xit}
\eeq
where $C_\ell^{TT}$ is the unlensed temperature power spectrum, $ \theta \equiv |\bm{\theta}| = |\mathbf{x} - \mathbf{x'}|$ is the distance 
between two points on the plane of the sky and
\beq
A_0(\theta)\equiv\frac{1}{2 \pi}\int \ell^3 \ d\ell \ C_\ell^{\psi \psi} J_0(\ell \theta);~A_2(\theta)\equiv\frac{1}{2 \pi}\int \ell^3\ d\ell
\  C_\ell^{ \psi \psi}J_2(\ell\theta);~\sigma^2(\theta)\equiv A_0(0)-A_0(\theta).
\eeq

The CMB polarization field can be expressed in terms of the Stokes parameters as
$P(\mathbf{x}) = Q(\mathbf{x}) + i U(\mathbf{x})$,
which can then be expanded in Fourier space in terms of the $E$ and $B$-modes:
\begin{equation}
 P(\mathbf{x}) = - \int \frac{d^2 \mathbf{l}}{2 \pi} [E(\mathbf{l}) + i B(\mathbf{l})] e^{-2 i \phi_\mathbf{l}} e^{ i \mathbf{l}
 \cdot \mathbf{x}}
\end{equation}
where $\phi_\mathbf{l}$ is the angle made by the vector $\mathbf{l}$ with the $x$-axis.
In the basis defined by the direction $\bm \theta = \bf{x} - \bf{x}'$,
the three lensed polarization correlation functions are given by (the subscript ``$r$'' denotes the quantities being measured in this basis):
\begin{eqnarray}
 \tilde \xi_+(\theta) &=& \langle P^*_r(\mathbf{x} + \bm{\alpha}) P_r(\mathbf {x}' + \bm{\alpha}') \rangle,\
 \tilde  \xi_-(\theta) = \langle P_r(\mathbf{x} + \bm{\alpha}) P_r(\mathbf{x}'+ \bm{\alpha}') \rangle, \
 \tilde \xi_\times(\theta) = \langle P_r(\mathbf{x}+ \bm{\alpha}) T(\mathbf{x}'+ \bm{\alpha}') \rangle.
\end{eqnarray}
Now, retaining terms upto the second order in $A_2(\theta)$, the above expressions can be computed to be \cite{Lewis}:
\bea
\tilde{\xi}_+(\theta)&=&\frac{1}{2\pi}\int \ell'd\ell' \left(C_{\ell'}^{EE}+C_{\ell'}^{BB}\right)e^{-\ell'^2
\sigma^2(\theta)/2}\left[\left(1+\frac{1}{16}\ell'^4A_2(\theta)^2\right)J_0(\ell'\theta)
+\frac{1}{2}\ell'^2 A_2(\theta)J_2(\ell'\theta)+\frac{1}{16}\ell'^4 A_2(\theta)^2J_4(\ell'\theta)\right]\label{corrp}\nonumber\\
\tilde{\xi}_-(\theta)&=&\frac{1}{2\pi}\int \ell'd\ell' \left(C_{\ell'}^{EE}-C_{\ell'}^{BB}\right)e^{-\ell'^2\sigma^2(\theta)/2}
\left[\left(1+\frac{1}{16}\ell'^4 A_2(\theta)^2\right)J_4(\ell'\theta)+\frac{1}{4}\ell'^2A_2(\theta)\left[J_2(\ell'\theta)+
J_6(\ell'\theta)\right]\right.\label{corrm}\nonumber\\
&+&\left.\frac{1}{32}\ell'^4A_2(\theta)^2\left[J_0(\ell'\theta)+J_8(\ell'\theta)\right]\right]\\
\tilde{\xi}_\times(\theta)&=&\frac{1}{2\pi}\int \ell'd\ell'C_{\ell'}^{TE}e^{-\ell'^2\sigma^2(\theta)/2}
\left[\left(1+\frac{1}{16}\ell'^4A_2(\theta)^2\right)J_2(\ell'\theta)+\frac{1}{4}\ell'^2A_2(\theta)
\left[J_0(\ell'\theta)+J_4(\ell'\theta)\right]\right.\label{corrc}\nonumber\\
&+&\left.\frac{1}{32}\ell'^4A_2(\theta)^2\left[J_2(\ell'\theta)+J_6(\ell'\theta)\right]\right]\nonumber.
\eea
Here $C_{\ell}^{TE}$, $C_{\ell}^{EE}$ and $C_{\ell}^{BB} $ are the unlensed cross-correlation spectra and unlensed $E$- and $B$-mode power spectra respectively. 
The unlensed correlation functions can be obtained from Eqns.\eqref{xit} and \eqref{corrp} by setting
$\sigma(\theta)=A_2(\theta)=0$. Once the correlation functions are known, the corresponding power spectra can be evaluated using the following
relations:
\bea
\tilde{C}_\ell^{TT}&=&2\pi\int \theta \ d \theta J_0(\ell \theta) \ \tilde{\xi}(\theta),~~{C}_\ell^{TT}=2\pi\int\theta \ d \theta J_0( \ell \theta) \ {\xi}(\theta)
\label{clt}\\
\tilde{C}_\ell^+\equiv\tilde{C}_l^{EE}+\tilde{C}_\ell^{BB} &=& 2\pi\int\theta \ d \theta J_0(\ell\theta)\ \tilde{\xi}_+(\theta),~~
{C}_\ell^+= 2\pi\int\theta \ d \theta J_0(\ell\theta) \ {\xi}_+(\theta)
\label{clb}\\
\tilde{C}_\ell^-\equiv\tilde{C}_\ell^{EE}-\tilde{C}_\ell^{BB} &=& 2\pi\int\theta \ d \theta J_4(\ell\theta) \ \tilde{\xi}_-(\theta),~~
{C}_\ell^-= 2\pi\int\theta \ d \theta J_4(\ell\theta) \ {\xi}_-(\theta)\label{cle}\\
\tilde{C}_\ell^{TE} &=& 2\pi\int\theta \ d \theta J_2(\ell\theta) \ \tilde{\xi}_\times(\theta),~~
{C}_\ell^{TE}= 2\pi\int\theta \ d \theta J_2(\ell\theta) \ {\xi}_\times(\theta).\label{clte}
\eea
For our investigation in the flat-sky approximation, we shall estimate the difference between the lensed and unlensed CMB spectra using Eqns. 
\eqref{clt}-\eqref{clte}.


\vspace{0.cm}
\subparagraph*{\bf Full-Sky Results:}
It can be shown that in case of the full-sky, the lensed temperature anisotropy correlation function, upto second order in $A_2 (\beta)$, 
is given by \cite{Lewis}:
\beq
\tilde{\xi}(\beta)\approx \sum_{\ell}\frac{2\ell+1}{4\pi}C_\ell^{TT}\left\{X^2_{000}(\beta)d^\ell_{00}(\beta)+\frac{8}{\ell(\ell+1)} 
A_2 (\beta)X'^2_{000}(\beta)d^{ \ell'}_{\ell -1}(\beta)+
 A_2^2(\beta)\left(X'^2_{000}d^\ell_{00}(\beta)+X^2_{220}d^\ell_{2-2}(\beta)\right)\right\} \, ,
\label{xitfull}
\eeq
where $\cos\beta = \hat{\bf n}_1 \cdot \hat{\bf n}_2$, $d^\ell_{mm'}$'s are the standard rotation matrices and 
now, in the full sky case, we have defined,
\bea
A_0(\beta)&=&\sum_{\ell}\frac{2\ell+1}{4\pi}\ell(\ell+1)C_\ell^{\psi \psi}d^\ell_{11}(\beta) ;\
A_2 (\beta)=\sum_{\ell}\frac{2\ell+1}{4\pi}\ell(\ell+1)C_\ell^{\psi \psi}d^\ell_{-11}(\beta);  \
\sigma^2(\beta)\equiv A_0(0)-A_0(\beta)
\eea
with
\beq
X_{imn}\equiv \int_0^{\infty}\frac{2\alpha}{\sigma^2(\beta)}\left(\frac{\alpha}{\sigma^2(\beta)}\right)^i e^{-\alpha^2/{\sigma^2(\beta)}}d^\ell_{mn}(\alpha)d\alpha
\eeq
and the prime denotes derivative with respect to $\sigma^2(\beta)$.

In the full-sky, the three correlation functions for the polarization can be expressed upto second order in $A_2 (\beta)$,
as follows \cite{Lewis}:
\bea
\tilde{\xi}_+(\beta)&\approx&\sum_{\ell}\frac{2\ell+1}{4\pi} \left(C_\ell^{EE}+C_\ell^{BB}\right)\left\{X^2_{022}d^\ell_{22}+2 A_2 (\beta) X_{132}
X_{121}d^\ell_{31}+  A_2 (\beta)^2\left[X^{'2}_{022} d^\ell_{22}+X_{242}X_{220}d^\ell_{40}\right]\right\} \label{xipfull}\\
\tilde{\xi}_-(\beta)&\approx&\sum_{\ell}\frac{2\ell+1}{4\pi} \left(C_\ell^{EE}-C_\ell^{BB}\right)
\left\{X^2_{022}d^\ell_{2-2}+ A_2 (\beta) \left[X^2_{121}d^\ell_{1-1}+
X^2_{132}d^\ell_{3-3}\right] \right.\nonumber \label{ximfull} \\
&+&\left.\frac{1}{2}  A_2 (\beta)^2\left[2X^{'2}_{022} d^\ell_{2-2}+X^2_{220}d^\ell_{00} + X^2_{242}d^\ell_{4 -4}\right]\right\}\\
\tilde{\xi}_\times(\beta)&\approx&\sum_{\ell}\frac{2\ell+1}{4\pi} C_\ell^{TE}
\left\{X_{022}X_{000} d^\ell_{02}+ A_2 (\beta) \left[\frac{2X'_{000}}{\sqrt{\ell(\ell+1)}}(X_{112}d^\ell_{11}+
X_{132}d^\ell_{3 -1})\right] \right.\nonumber \\
&+&\left.\frac{1}{2}  A_2 (\beta)^2\left[(2X^{'}_{022}X'_{000} + X^2_{220})d^\ell_{20} + X_{220}X_{242}d^\ell_{-24}  \right]\right\}\label{xicfull}
\eea
Again, the power spectra are related to the correlation functions by the following equations for the full-sky:
\begin{eqnarray}
\tilde{C}_\ell^{TT} &=& 2\pi\int_{-1}^1 \tilde{\xi}(\beta) d^\ell_{00}(\beta) d(\cos \beta) \ ; \ {C}_\ell^{TT} = 2\pi\int_{-1}^1 {\xi}(\beta) 
d^\ell_{00}(\beta) d(\cos \beta)
\label{clfulltt} \\
\tilde{C}_\ell^{EE} +  \tilde{C}_\ell^{BB} \equiv \tilde{C}_\ell^{+} &=& 2\pi\int_{-1}^1 \tilde{\xi}_{+}(\beta) d^\ell_{22}(\beta) d(\cos \beta)\ ; 
\ {C}_\ell^{+} = 2\pi\int_{-1}^1 {\xi}_{+}(\beta) d^\ell_{22}(\beta) d(\cos \beta)  \label{clfullp} \\
\tilde{C}_\ell^{EE} - \tilde{C}_\ell^{BB} \equiv \tilde{C}_\ell^{-} &=& 2\pi\int_{-1}^1 \tilde{\xi}_{-}(\beta) d^\ell_{2 -2}(\beta) d(\cos \beta) \ ; 
\ {C}_\ell^{-} = 2\pi\int_{-1}^1 {\xi}_{-}(\beta) d^\ell_{2 -2}(\beta) d(\cos \beta) \label{clfullm} \\
\tilde{C}_\ell^{TE} &=& 2\pi\int_{-1}^1 \tilde{\xi}_\times(\beta) d^\ell_{20}(\beta) d(\cos \beta) \ ; 
\ {C}_\ell^{TE} = 2\pi\int_{-1}^1 {\xi}_\times(\beta) d^\ell_{20}(\beta) d(\cos \beta)
\label{clfullte}
\end{eqnarray}
The above Eqns. \eqref{clfulltt}, \eqref{clfullp}, \eqref{clfullm}, \eqref{clfullte} are used in our full-sky analysis for estimating the unlensed CMB spectra. 

\section{The unlensed CMB spectra using matrix inversion technique}\label{mit}
Following the discussions in the last section, let us now address the main objective of our paper, i.e., to find a method for extracting
the \textit{unlensed intrinsic CMB spectra} from the lensed ones under ideal conditions. We show that a simple 
\textit{matrix inversion technique} can be utilized to subtract the lensing contribution with very good accuracy.
Before going into the technical details, we note that,
 since the intrinsic CMB spectra are very important in the context of present day cosmology, especially as regards the primordial gravitational
waves, it is important to subtract the lensing contribution.

\subsection{\bf Flat Sky Analysis} 

In the flat-sky limit, we solve Eqns.(\ref{clt})-(\ref{clte}) for the unlensed power spectra.
The lensed CMB power spectra can be estimated directly from the observations. These can then be utilized to obtain
the corresponding unlensed power spectra provided we have the lensing potential, using matrix inversion technique as follows.

Using Eqn. \eqref{clt}, we first calculate the difference between the lensed and the unlensed CMB temperature anisotropy
power spectrum, which can be expressed as:
\beq
\tilde{C}_\ell^{TT}-{C}_\ell^{TT} = 2\pi\int \theta \ d  \theta J_0(\ell \theta) (\tilde{\xi}(\theta) - {\xi}(\theta)).
\label{cltt}
\eeq
With the help of Eq.\eqref{xit},  the above expression can be rewritten in the following form:
\bea
\tilde{C}_\ell^{TT}-{C}_\ell^{TT} &=& \int \theta \ d  \theta J_0(\ell \theta) \int d\ell'\ \ell'\ C_{\ell'}^{TT}e^{-\ell'^2\sigma^2(\theta)/2}
\left[\left(1+\frac{1}{16}\ell'^4 A_2(\theta)^2\right)J_0( \ell' \theta)
+\frac{1}{2}\ell'^2 A_2(\theta)J_2(\ell'\theta)+\frac{1}{16}\ell'^4 A_2^2 J_4(\ell'\theta )\right] \nonumber \\
&-& \int \theta \ d  \theta J_0(\ell \theta) \int d\ell'\ \ell' \ C_{\ell'}^{TT} J_0( \ell' \theta) \nonumber \\
&\equiv&  \int d\ell' \ C_{\ell'}^{TT} \ \delta k^T(\ell,\ell')\label{difftt}.
\eea
In the last step, we have changed the order of integration and have defined the \textit{temperature kernel} $\delta k^T(\ell,\ell')$, which can be read off from the above equation as
\beq
\delta k^T(\ell,\ell')\equiv \ell'\int\theta \ d \theta  J_0(l\theta)\left\{e^{-\ell'^2\sigma^2(\theta)/2}\left[\left(1+\frac{\ell'^4}{16}A_2(\theta)^2\right)J_0(\ell'\theta)+\frac{\ell'^2}{2}A_2(\theta)
J_2(\ell'\theta)+\frac{\ell'^4}{16}A_2(\theta)^2J_4(\ell'\theta)\right]-J_0(\ell'\theta)\right\}.
\eeq
Eqn.\eqref{difftt} is the  \textit{Fredholm 
integral equation of the second kind}. Generally, these equations are very difficult to solve both analytically and numerically. 
We now provide a  technique to solve this 
equation numerically that works satisfactorily well whenever the kernel functions are small, which is the relevant case here, as the lensing contributions are small.
To do this, we rewrite Eqn.\eqref{difftt} in the following form:
\begin{equation}
\tilde{C}_\ell^{TT} = \int d\ell' \ C_{\ell'}^{TT}\left[\delta(\ell - \ell')+ \delta k^T(\ell,\ell')\right]
\approx\sum_{\ell'} C_{\ell'}^{TT}\left[I_{\ell\ell'}+\delta k^T_{\ell\ell'}\right].
\end{equation}
In the last step, we have discretized the continuous expression by replacing the integral over $\ell'$ by a sum over the various $\ell'$,
and the delta function by its discrete counterpart, the identity matrix $I_{\ell\ell'}$. $\delta k^T_{\ell\ell'}$ is the discrete-$\ell$ representation of
$\delta k^T(\ell,\ell')$. Note that the superscript ``$T$'' of $\delta k^T_{\ell\ell'}$ denotes temperature.
 After defining $M^T_{\ell\ell'}\equiv I_{\ell\ell'}+\delta k^T_{\ell\ell'}$, we can rewrite the above expression for the lensed temperature angular 
 power spectrum as a set of linear
equations, which, in matrix notation has the form:
\begin{equation}
{\bf\tilde{C}^{TT}}={\bf M^TC^{TT}}.
\label{ulclt}
\end{equation}
As the elements of the kernel matrix 
$\delta k^T(\ell,\ell')$ are very small, we can solve Eqn.\eqref{ulclt} by 
expanding $(\mathbf{M^T})^{-1}$ as a Taylor series in powers of $\mathbf{\delta k^T}$. Hence, Eqn.\eqref{ulclt} may be rewritten as:
\bea
\mathbf{C^{TT}}&=& \mathbf{(M^{T})^{-1}}\mathbf{\tilde{C}^{TT}}\nonumber\\
  &=& {\bf\left[I-\delta k^T+(\delta k^T)^2+\mathcal{O}((\delta k^T)^3)\right]\tilde{C}^{TT}}\label{ultem}
\eea
For our analysis, we have retained terms upto the second order in $\mathbf{\delta k^T}$.
Since the elements of $\mathbf{\delta k^T}$ are very small, $\mathbf{M^T}$ is very close to the identity, as a result, Eqn.\eqref{ulclt} can
also be solved exactly by taking the inverse of $\mathbf{M^T}$. This enables us to extract the unlensed temperature power spectrum $\mathbf{C^{TT}}$ from
the lensed one in a simple manner.

We now consider the case of the polarization spectra. Here, the relevant expressions are Eqns. (\ref{clb}), (\ref{cle}) and (\ref{clte}),
together with the expressions for the correlation functions, Eqns. (\ref{corrp}). We can follow the above procedure for these equations
to define the corresponding kernels, $\delta k^+$, $\delta k^-$ and $\delta k^\times$, in terms of which the lensed CMB polarization and cross
power spectra can be expressed as:
\begin{eqnarray}
\tilde{C}_\ell^{+} &=& \int d\ell' \ C_{\ell'}^{+}\left[\delta(\ell - \ell')+ \delta k^+(\ell,\ell')\right]
\approx\sum_{\ell'} C_{\ell'}^{+}\left[I_{\ell\ell'}+\delta k^+_{\ell\ell'}\right] \nonumber \\
\tilde{C}_\ell^{-} &=& \int d\ell' \ C_{\ell'}^{-}\left[\delta(\ell - \ell')+ \delta k^-(\ell,\ell')\right]
\approx\sum_{\ell'} C_{\ell'}^{-}\left[I_{\ell\ell'}+\delta k^-_{\ell\ell'}\right]\nonumber \\
\tilde{C}_\ell^{TE} &=& \int d\ell' \ C_{\ell'}^{TE}\left[\delta(\ell - \ell')+ \delta k^\times(\ell,\ell')\right]
\approx\sum_{\ell'} C_{\ell'}^{TE}\left[I_{\ell\ell'}+\delta k^\times_{\ell\ell'}\right]
\end{eqnarray}
where we now have:
\bea
\delta k^+(\ell,\ell')&\equiv& \ell'\int \theta \ d\theta  J_0(\ell\theta)\left\{e^{-\ell'^2\sigma^2(\theta)/2}\left[\left(1+\frac{\ell'^4}{16}A_2(\theta)^2\right)J_0(\ell'\theta)+\frac{\ell'^2}{2}A_2(\theta)
J_2(\ell'\theta)+\frac{\ell'^4}{16}A_2(\theta)^2J_4(\ell'\theta)\right]-J_0( \ell'\theta)\right\}\nonumber\\
\delta k^-(\ell,\ell')&\equiv& \ell'\int  \theta \ d \theta   J_4(l \theta )\left\{e^{-\ell'^2\sigma^2( \theta )/2}\left[\left(1+\frac{1}{16}\ell'^4 A_2(\theta) ^2\right)J_4(\ell' \theta )+\frac{1}{4}\ell'^2 A_2(\theta) \left[J_2(\ell' \theta )+J_6(\ell' \theta )
\right]\right.\right.\nonumber\\
&+&\left.\left.\frac{1}{32}\ell'^4 A_2(\theta) ^2\left[J_0(\ell' \theta )+J_8(\ell' \theta )\right]\right]-J_4(\ell'  \theta )\right\}\label{kulminus}\\
\delta k^\times(\ell,\ell')&\equiv& \ell'\int\theta \ d \theta  J_2(l \theta )\left\{e^{-\ell'^2\sigma^2( \theta )/2}\left[\left(1+\frac{1}{16}\ell'^4A_2(\theta)^2\right)J_2(\ell' \theta )+\frac{1}{4}\ell'^2 A_2(\theta) \left[J_0(\ell' \theta )
+J_4(\ell' \theta )\right]\right.\right.\nonumber\\
&+&\left.\left.\frac{1}{32}\ell'^4 A_2(\theta) ^2\left[J_2(\ell' \theta )+J_6(\ell' \theta )\right]\right]-J_2(\ell' \theta)\right\}\label{kulcross}\nonumber
\eea
Analogous to the case of temperature power spectra, we can also define the corresponding kernel matrices for the polarization and the cross power spectra,
through relations similar to that in Eqn. (\ref{ultem}). This can be done as follows:
\bea
{\bf C^+}&=& {\bf\left[I-\delta k^++(\delta k^+)^2+\mathcal{O}((\delta k^+)^3)\right]\tilde{C}^+}\label{ulplus}\\
{\bf C^-}&=& {\bf\left[I-\delta k^-+(\delta k^-)^2+\mathcal{O}((\delta k^-)^3)\right]\tilde{C}^-}\label{ulminus}\\
{\bf{C}^{TE}}&=&{\bf \left[I-\delta k^\times+(\delta k^\times)^2+\mathcal{O}((\delta k^\times)^3)\right]\tilde{C}^{TE}}\label{ulcross}
\eea
The unlensed power spectra are obtained by solving the above Eqns. \eqref{ulplus}, \eqref{ulminus} and \eqref{ulcross}; we have:
\bea
\bf C^{EE}&=& \frac{1}{2}\left(\bf C^++C^-\right)\\
\bf C^{BB}&=& \frac{1}{2}\left(\bf C^+-C^-\right)\label{bb}
\eea

Thus, we see that with an estimate of the power spectrum for the lensing potential, it is possible, in principle, to extract the corresponding unlensed power spectra from the
lensed ones in a very simple manner. As a result, using our estimate, it may be possible to constrain the primordial
$B$-mode spectra once we have the lensed $B$-mode spectrum. In the realistic situation, our formulation must be convolved with estimates for the noise in the measured spectra and uncertainties in the transfer function; the present formalism serves as a demonstration of the deconvolution of the lensing effect under ideal conditions.

\vspace{0.cm}
\subsection{Full Sky Analysis}

We now repeat the above procedure for the case of the full-sky, the only difference being that the kernel functions are now different.
Using Eqns. \eqref{xitfull} and \eqref{xipfull}-\eqref{clfullte}, we can first
express the deviation of the lensed spectra from and unlensed ones; for the temperature anisotropy this reads:
\begin{eqnarray}
\tilde{C}_\ell^{TT} - C_\ell^{TT}
&=&  \sum_{\ell'} \frac{2 \ell' + 1}{2} \int_{0}^{\pi}   \sin \beta \ d \beta \ d^{\ell}_{00}(\beta) \ C_{\ell'}^{TT}  \Big\{X^2_{000}(\beta)d^{\ell'}_{00}(\beta)+
\frac{8}{\ell'(\ell'+1)} A_2 (\beta)X'^2_{000}(\beta)d^{ \ell'}_{1 -1}(\beta) \nonumber \\
&&+  A_2^2(\beta)\left(X'^2_{000}d^{\ell'}_{00}(\beta)+X^2_{220}d^{\ell'}_{2-2}(\beta)\right)\Big\}
 - \sum_{\ell'} \frac{2 \ell' + 1}{2} \int_{0}^{\pi}   \sin  \beta \ d \beta \ C_{\ell'}^{TT} d^{\ell}_{00}(\beta) d^{\ell'}_{00}(\beta)\nonumber\\
&\equiv&  \sum_{\ell'} C_{\ell'}^{TT} \delta K^T(\ell, \ell').
\label{clttexpand}
\end{eqnarray}
In the last line, we have defined the temperature kernel for the full sky as:
\bea
\delta K^T(\ell,\ell')&=&\frac{2\ell'+1}{2}\int^\pi_{0} \sin \beta \ d \beta \ d^\ell_{00}(\beta)\left\{\left[X^2_{000}-1\right]d^{\ell'}_{00}(\beta)+
\frac{8}{\ell'(\ell'+1)} A_2 (\beta) X'^2_{000}d^{\ell'}_{1 -1}(\beta) \right.\nonumber\\
&+&\left.
 A_2 (\beta) ^2\left(X'^2_{000}d^{\ell'}_{00}(\beta)+X^2_{220}d^{\ell'}_{2-2}(\beta)\right)\right\}
\eea
Similarly, for the case of the polarization power spectra, the corresponding lensed power spectra turn out to be:
\begin{eqnarray}
\tilde{C}_\ell^{+} &=& \sum_{\ell'} C_{\ell'}^{+}\left[I_{\ell\ell'}+\delta K^+(\ell,\ell')\right] \nonumber \\
\tilde{C}_\ell^{-} &=& \sum_{\ell'} C_{\ell'}^{-}\left[I_{\ell\ell'}+\delta K^-(\ell,\ell')\right]\nonumber \\
\tilde{C}_\ell^{TE} &=& \sum_{\ell'} C_{\ell'}^{TE}\left[I_{\ell\ell'}+\delta K^\times(\ell,\ell')\right]
\end{eqnarray}
where we have now defined
\bea
\delta K^+(\ell,\ell')&=&\frac{2\ell'+1}{2} \int^\pi_{0} \sin \beta \ d \beta  \  d^\ell_{22}(\beta)
\left\{\left[X^2_{022}-1\right]d^{\ell'}_{22}(\beta)+2 A_2 (\beta) X_{132}
X_{121}d^{\ell'}_{31}(\beta)\right.\nonumber\\
&+&\left.  A_2 (\beta)^2\left[X^{'2}_{022} d^{\ell'}_{22}(\beta)+X_{242}X_{220}d^{\ell'}_{40}(\beta)\right]\right\}\\
\delta K^-(\ell,\ell')&=& \frac{2\ell'+1}{2} \int^\pi_{0} \sin \beta \ d \beta  \  d^\ell_{2-2}(\beta)\left\{\left[X^2_{022}-1\right]d^{\ell'}
_{2-2}(\beta)+  A_2 (\beta) \left[X^2_{121}d^{\ell'}_{1-1}(\beta)+
X^2_{132}d^{\ell'}_{3-3}(\beta)\right] \right.\nonumber\\
&+&\left.\frac{1}{2}  A_2 (\beta)^2\left[2X^{'2}_{022} d^{\ell'}_{2-2}(\beta) + X^2_{220}d^{\ell'}_{00}(\beta) +X^2_{242}d^{\ell'}_{4 -4}(\beta)\right]\right\} \\
\delta K^\times(\ell,\ell')&=&\frac{2\ell'+1}{2} \int^\pi_{0} \sin \beta \ d \beta  \  d^{\ell'}_{20}(\beta)
\left\{\left[X_{022}X_{000}-1\right]d^{\ell'}_{20}(\beta)+  A_2 (\beta) \left[\frac{2X'_{000}}{\sqrt{\ell'(\ell'+1)}}(X_{112}d^{\ell'}_{11}(\beta)+
X_{132}d^{\ell'}_{3 -1}(\beta))\right] \right.\nonumber\\
&+&\left.\frac{1}{2}  A_2 (\beta)^2\left[(2X^{'}_{022}X'_{000} + X^2_{220})d^{\ell'}_{20}(\beta) + X_{220}X_{242}d^{\ell'}_{-24}(\beta)  \right]\right\}
\eea
In order to find the unlensed spectra, we again define kernel matrices for the temperature as well as for the polarization power spectra. Again, to order $(\delta \mathbf{K})^2$, 
the expressions for the unlensed spectra are given by:
\bea
\mathbf{C^{TT}}&=& {\bf\left[I-\delta K^T+(\delta K^T)^2+\mathcal{O}((\delta K^T)^3)\right]\mathbf{\tilde{C}}^{TT}}\label{sultem}\\
\mathbf{C^+}&=& {\bf\left[I-\delta K^++(\delta K^+)^2+\mathcal{O}((\delta K^+)^3)\right]\mathbf{\tilde{C}}^+}\label{sulplus}\\
\mathbf{C^-}&=& {\bf\left[I-\delta K^-+(\delta K^-)^2+\mathcal{O}((\delta K^-)^3)\right]\mathbf{\tilde{C}}^-}\label{sulminus}\\
\mathbf{C^{TE}}&=&{\bf \left[I-\delta K^\times+(\delta K^\times)^2+\mathcal{O}((\delta K^\times)^3)\right]\mathbf{\tilde{C}}^{TE}}
\label{sulcross}
\eea
In the above expressions, the $\mathbf{\delta K}$'s are the matrices associated with the kernels in the full sky.
Given the lensing power spectrum (which determines the $X_{imn}$ functions as well as $A_2 (\beta) $), the kernel functions can be worked out and using these kernels, the unlensed CMB power spectra are acquired by solving Eqns.\eqref{sultem}-\eqref{sulcross}. 
Hence, the unlensed CMB spectra can be extracted by subtracting the lensing artifact using the above method.

\section{Calibration, results and discussion}\label{num}

In this section, we describe a calibration of our  methodology against the unlensed spectra estimated from the 
primordial power spectra using 
CAMB. We also describe the numerical results obtained by applying the procedure described above to the lensed spectra taken from the best-fit WMAP 7-year
data (as mentioned, this is the latest available lensing data as of now), to calculate
the corresponding delensed spectra assuming zero unlensed $B$-mode power.

For the calibration of our technique, we start from the unlensed spectra generated by CAMB with the best-fit WMAP 7-year cosmological parameters, and use the lensing potential of CAMB to generate corresponding lensed 
spectra. We delens the lensed spectra thus obtained using our technique described earlier. Having obtained the 
delensed spectra, we calibrate them against
the original unlensed spectra. For the comparison, we calculate different kernel matrices, first adopting the flat-sky approximation and then 
considering  the sky to be spherical, using a FORTRAN90 code based on the lensing module of CAMB. The code requires as input the lensing power spectrum and 
generates the kernel matrices. We then utilize those matrices to obtain the delensed quantities from the corresponding lensed CMB spectra. For our investigation 
we have constructed  $2000\times2000$ kernel matrices, and the kernels are calculated explicitly for each integer value 
of $\ell$ (the multipole), without employing any interpolation.

\begin{figure}
\begin{center}
\includegraphics[width=8.5cm, height=5.5cm]{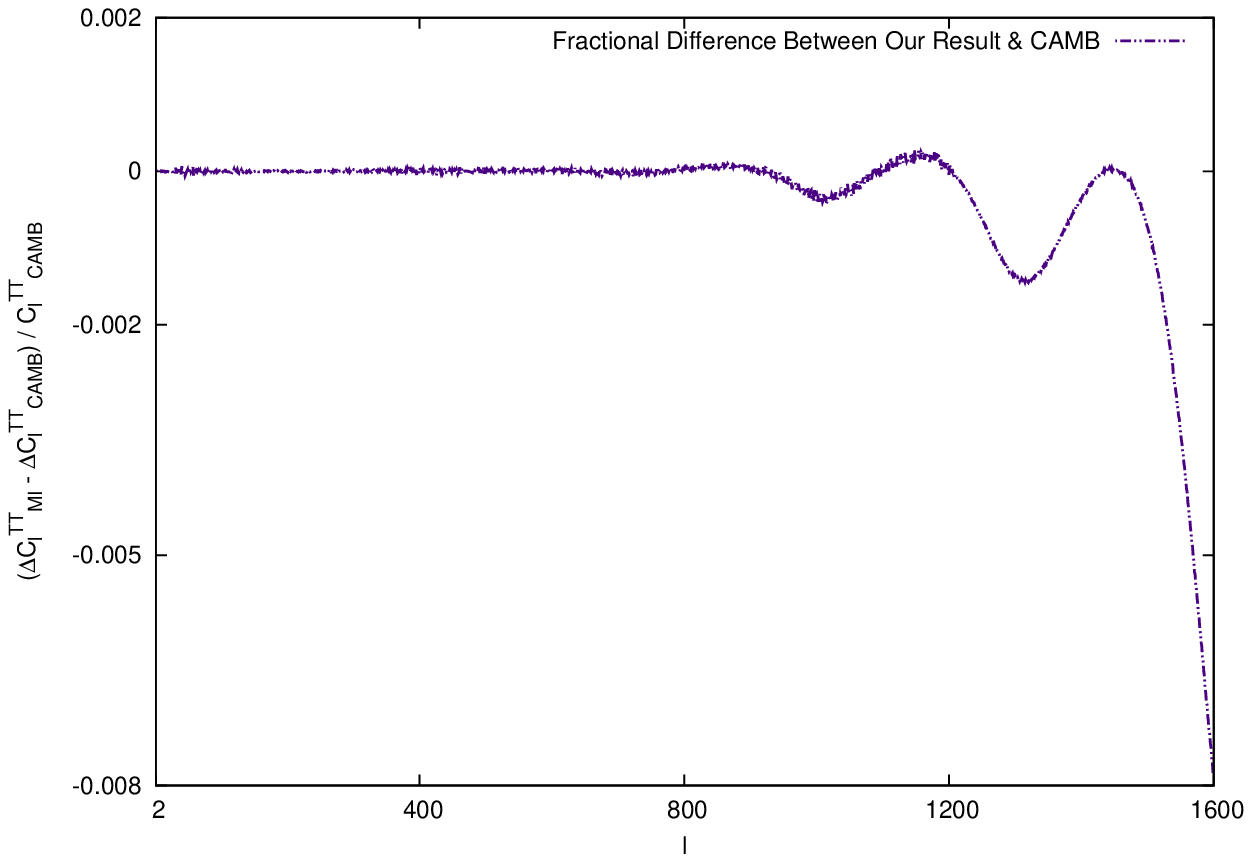}
\includegraphics[width=8.5cm, height=5.5cm]{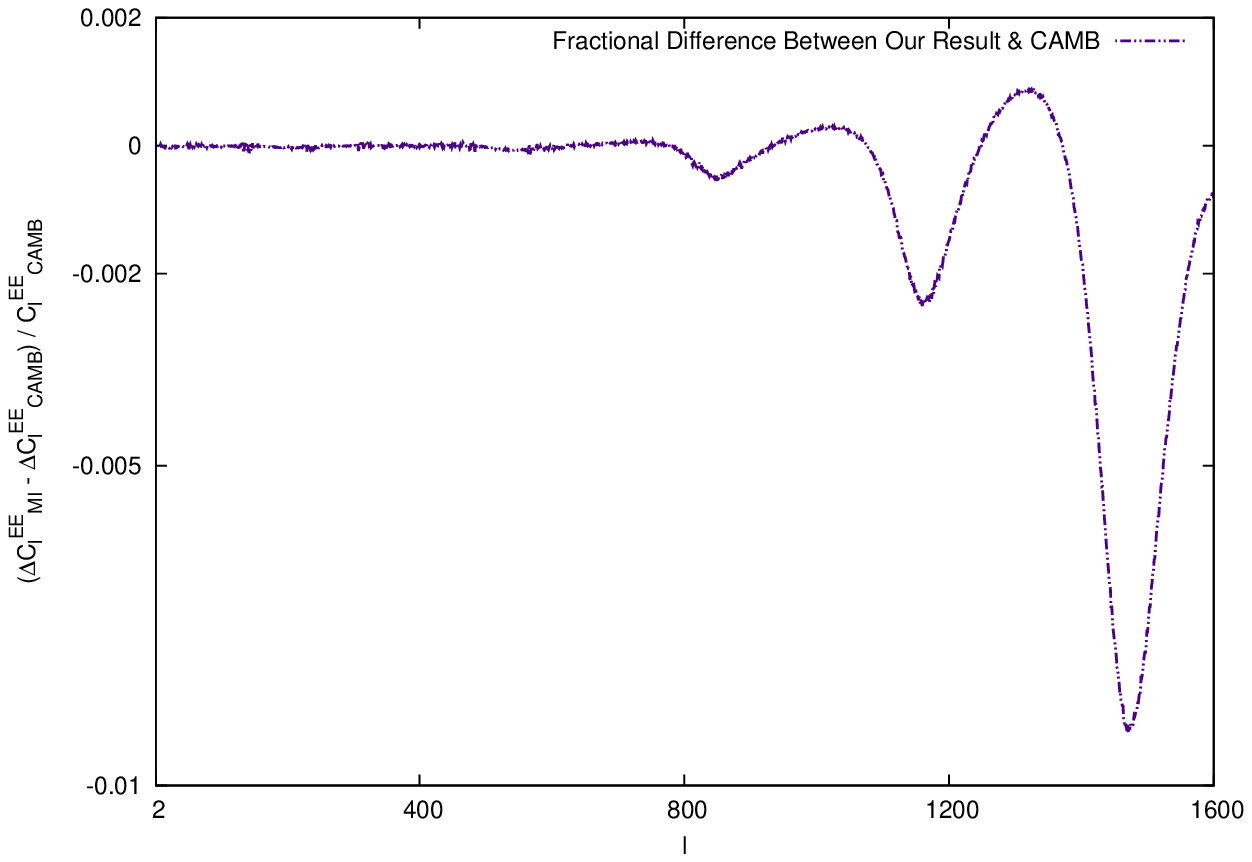}
\end{center}
\caption{The fractional difference between the lensed and delensed $TT$ (left panel) and $EE$ (right panel) power spectra   
as obtained from our procedure compared to the corresponding difference between the lensed and unlensed spectra
 calculated from CAMB.}
\label{fig:dttee}
\end{figure}

\begin{figure}
\begin{center}
\includegraphics[scale = 0.65]{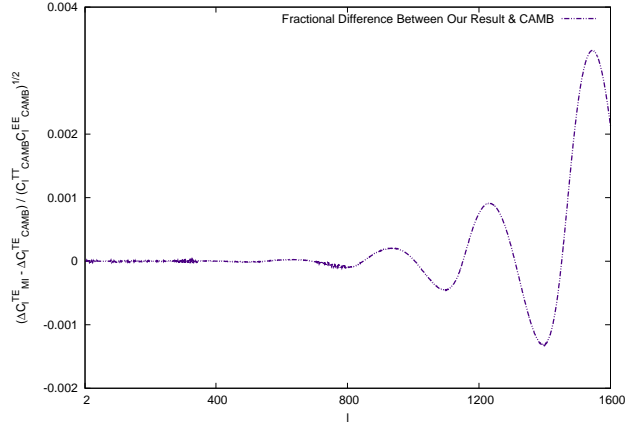}
\end{center}
\caption{The fractional difference between the lensed and delensed $TE$  power spectra  as obtained from our procedure compared to the corresponding difference between the lensed and unlensed spectra
calculated from CAMB. 
}
\label{fig:dte}
\end{figure}

In Figs.\ref{fig:dttee} and \ref{fig:dte}, we have plotted the fractional difference between the delensed spectra as obtained from our full-sky analysis, 
and the initial CAMB produced unlensed spectra started with originally. It can be seen that the differences are very small, of the order of $10^{-4}$ for values of 
$\ell \lesssim 1000$, for each of the $TT$, 
$TE$ and polarization spectra. For higher multipoles, the variation increases and reaches about 
$0.8\%$ for $TT$ and about 
$0.92\%$ for $EE$, and $0.33\%$ for $TE$  spectra between $\ell \sim 1500$-$1600$. This serves as a calibration for 
the accuracy of our procedure. The spectra obtained by CAMB are calculated 
independently using the primordial power spectrum together with Eqn.\eqref{pps}, and we recover the spectra 
started out with, to an accuracy of $0.33\% -0.92\%$ around $\ell \sim 1500$-$1600$. The deviation from the CAMB 
unlensed spectra is anticipated as we have proceeded through the reverse route by using the lensed CAMB spectra as 
input. Further, if it is possible to reconstruct the lensing potential power spectra directly from the observational data, 
independent of any model, then, by following our above procedure, the intrinsic CMB power may also be reconstructed in  a model independent 
manner. As an additional calibration, we have repeated the above procedure using the WMAP 9-year cosmological parameters, and we again find an identical accuracy of about 
$0.8\%$ for $TT$, about 
$0.92\%$ for $EE$, and $0.33\%$ for $TE$  spectra around $\ell \sim 1500$-$1600$ with these parameters, 
demonstrating that the accuracy of the technique is stable to variation in the input parameters.

\begin{figure}
\begin{center}
\includegraphics[width=8.5cm, height=5.5cm]{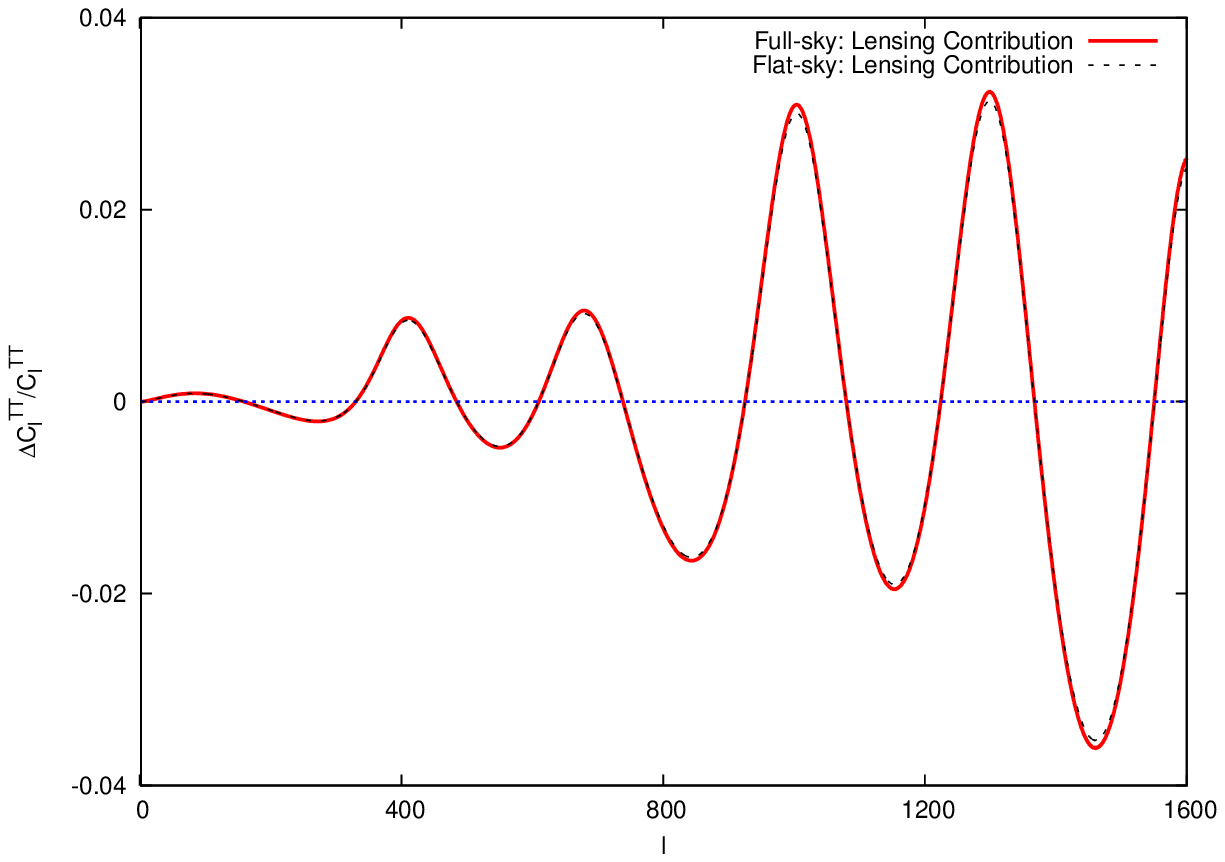}
\includegraphics[width=8.5cm, height=5.5cm]{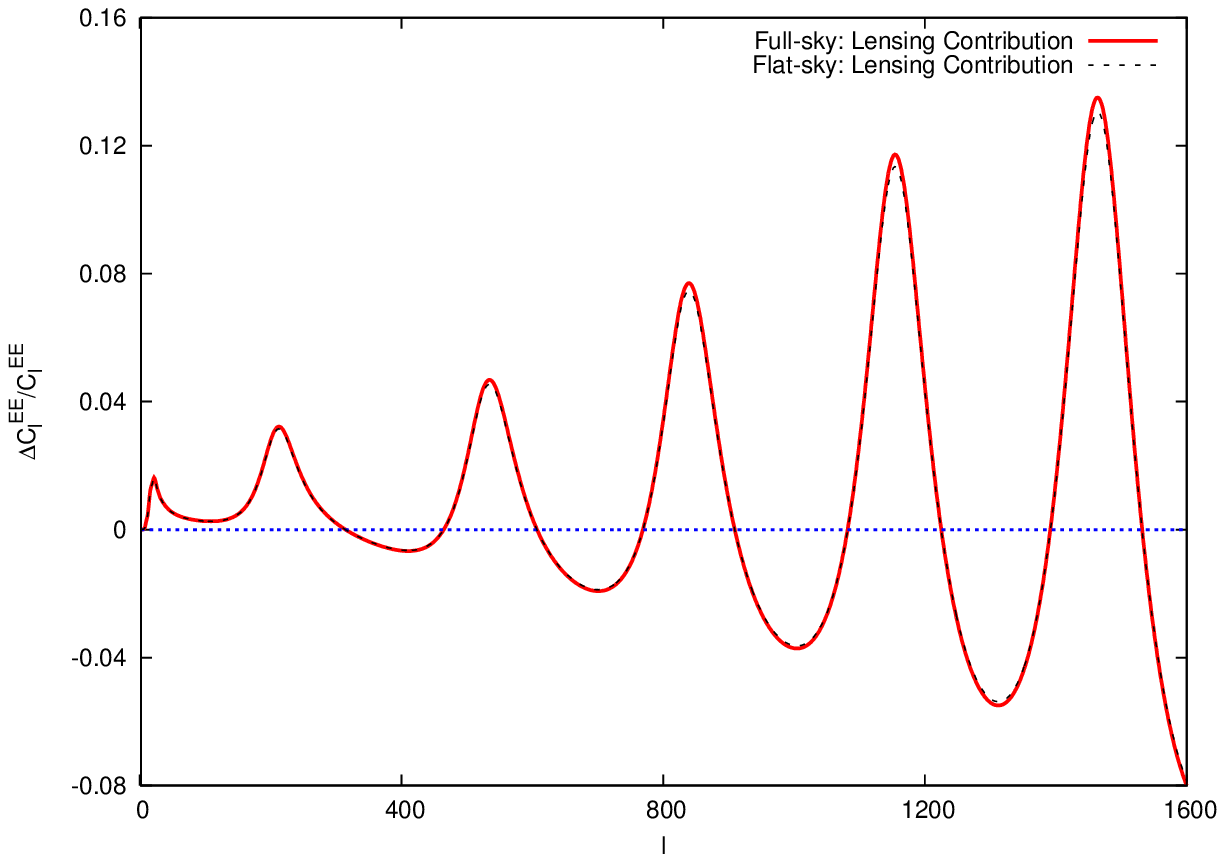}
\end{center}
\caption{This shows the fractional difference between the lensed and delensed $TT$ and $EE$ power spectra as obtained by our above 
procedure in the flat-sky approximation (shown by black dashed lines), and in the full-sky (shown by red solid lines). The lensed spectra used are 
from the WMAP 7-year best-fit $\Lambda CDM$ + TENS data. Hence, the above plots describe the \textit{lensing contributions}, in the flat-sky limit and in the full 
sky. It can be seen that the difference is very small, for this range of $\ell$-values.}
\label{fig:fcttee}
\end{figure}

\begin{figure}
\begin{center}
\includegraphics[width=8.5cm, height=5.5cm]{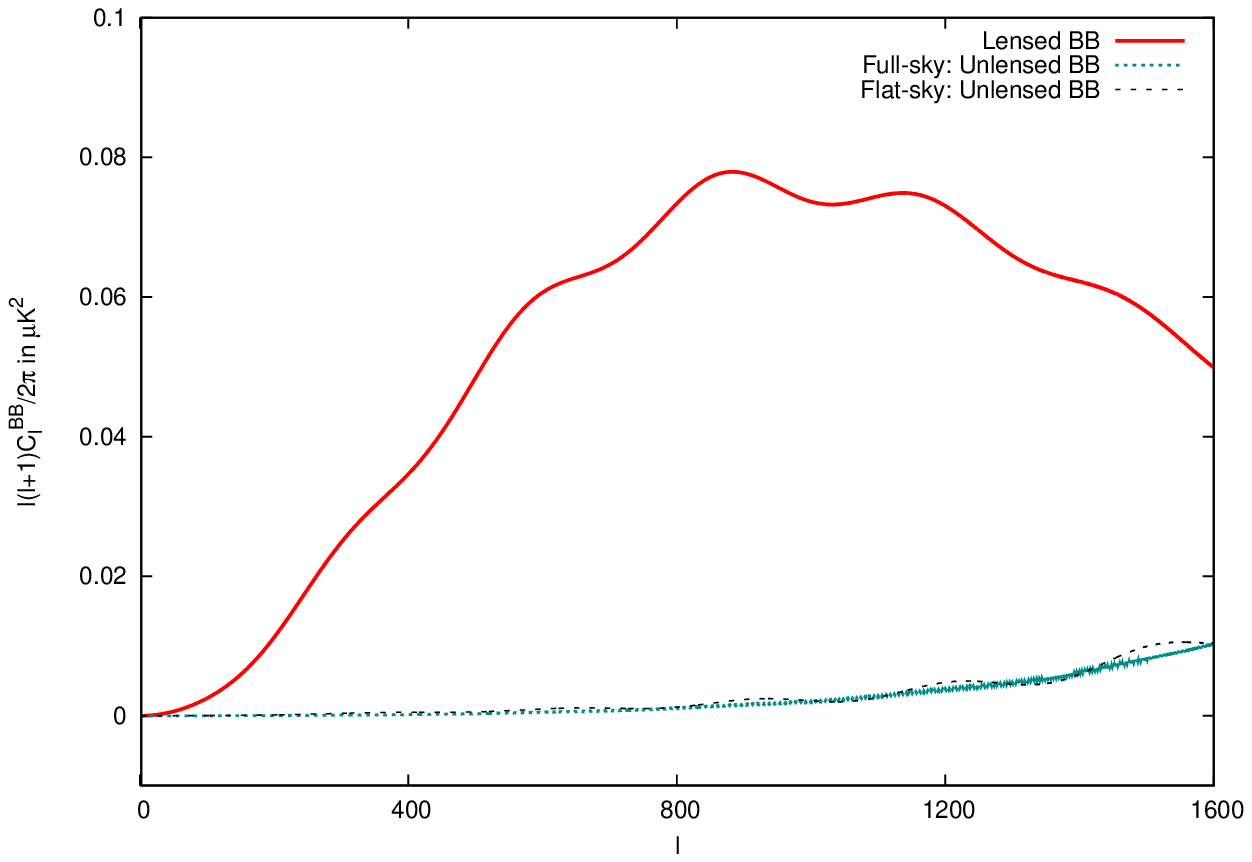}
\includegraphics[width=8.5cm, height=5.5cm]{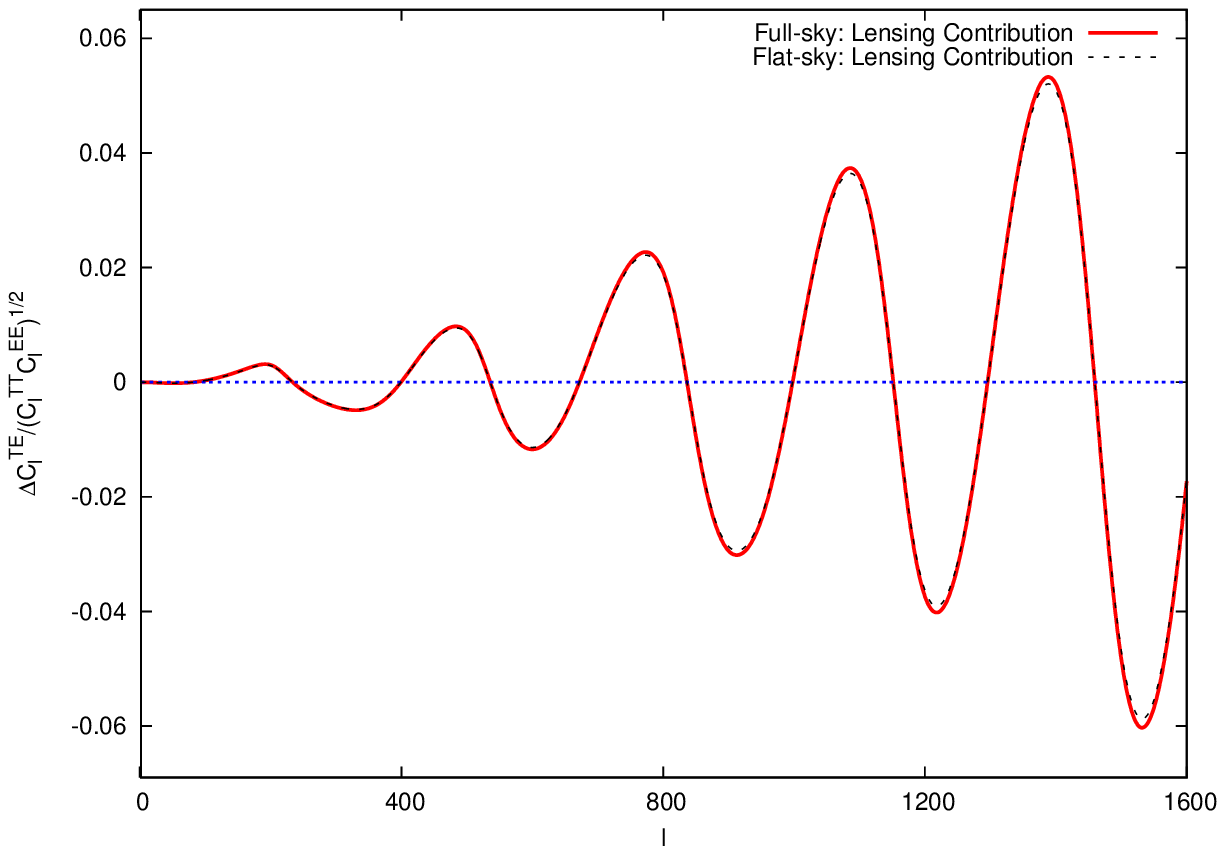}
\end{center}
\caption{The left panel shows the delensed $BB$ power spectrum as obtained by our above procedure in the flat-sky approximation (black dashed line), 
and in the full-sky (blue dot-dashed line). The lensed $BB$ spectrum (red solid line) from the WMAP 7-year best-fit data, is also plotted to provide a comparison.  
The right panel shows the lensing contribution to the $TE$ power spectrum  in the flat-sky approximation (black dashed lines), and in the full-sky (red solid lines).
 Again, it can be seen that the difference is very small, for this range of $\ell$-values.}
\label{fig:fcbbte}
\end{figure}

We now present results for the delensing of the WMAP7 spectra by our analysis, done in both the flat-sky approximation as well as in the spherical sky. For this, we have utilized the lensed spectra obtained from WMAP 7-year best-fit $\Lambda CDM$ + TENS data, assuming zero unlensed $B$-mode power, and delensed the spectra using the matrix inversion technique with the kernel matrices generated using the WMAP7 best-fit lensing potential. 
In Figs.\ref{fig:fcttee} and \ref{fig:fcbbte}, we have plotted the lensing contributions as estimated from our analysis for both the flat-sky and the full-sky. For
comparison, we have assumed zero intrinsic $B$-mode power. For values of $\ell \lesssim 1600$, we find that the two methods provide consistent 
results. The difference between the flat-sky and the full-sky results is very small in the large scale regime, and increases very slowly as we go to smaller scales, however, the 
fractional difference always remains less than $10^{-3}$.
 

\begin{figure}
\begin{center}
\includegraphics[width=8.5cm, height=5.5cm]{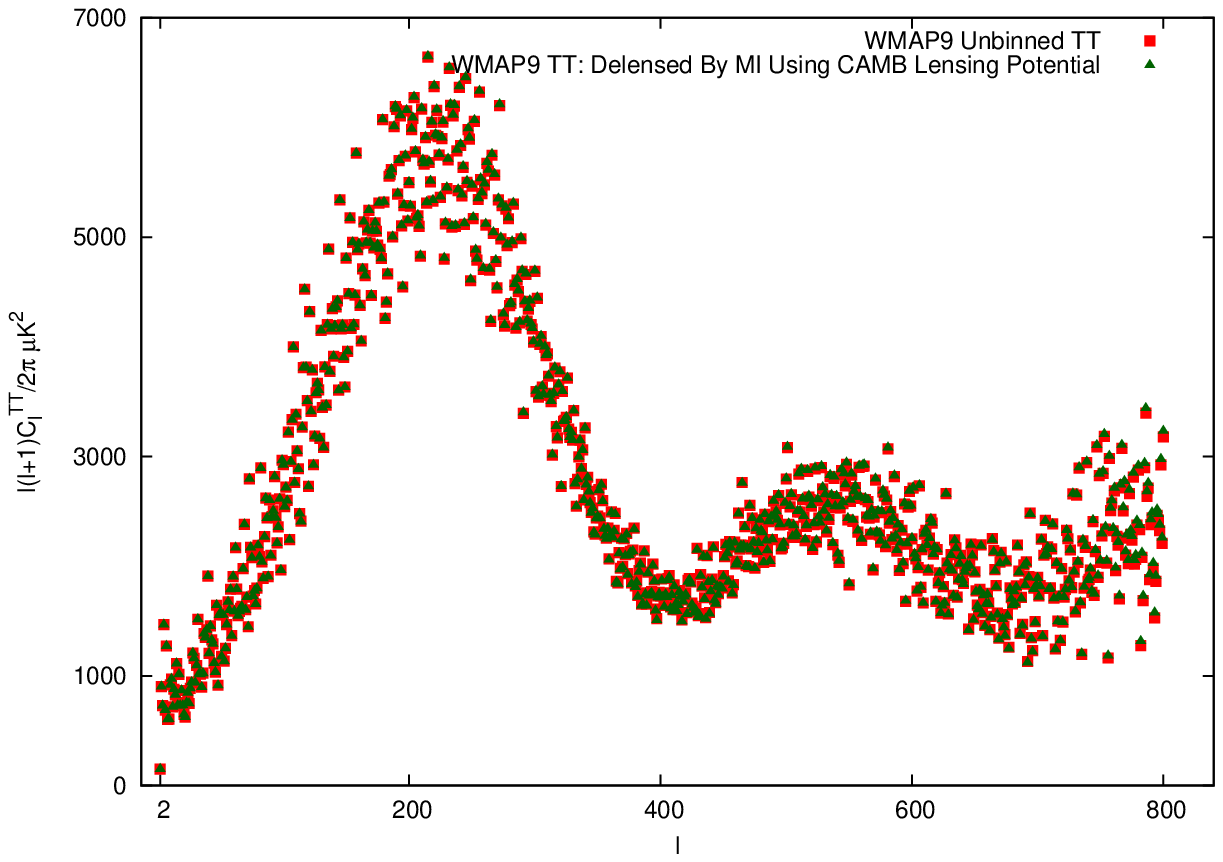}
\includegraphics[width=8.5cm, height=5.5cm]{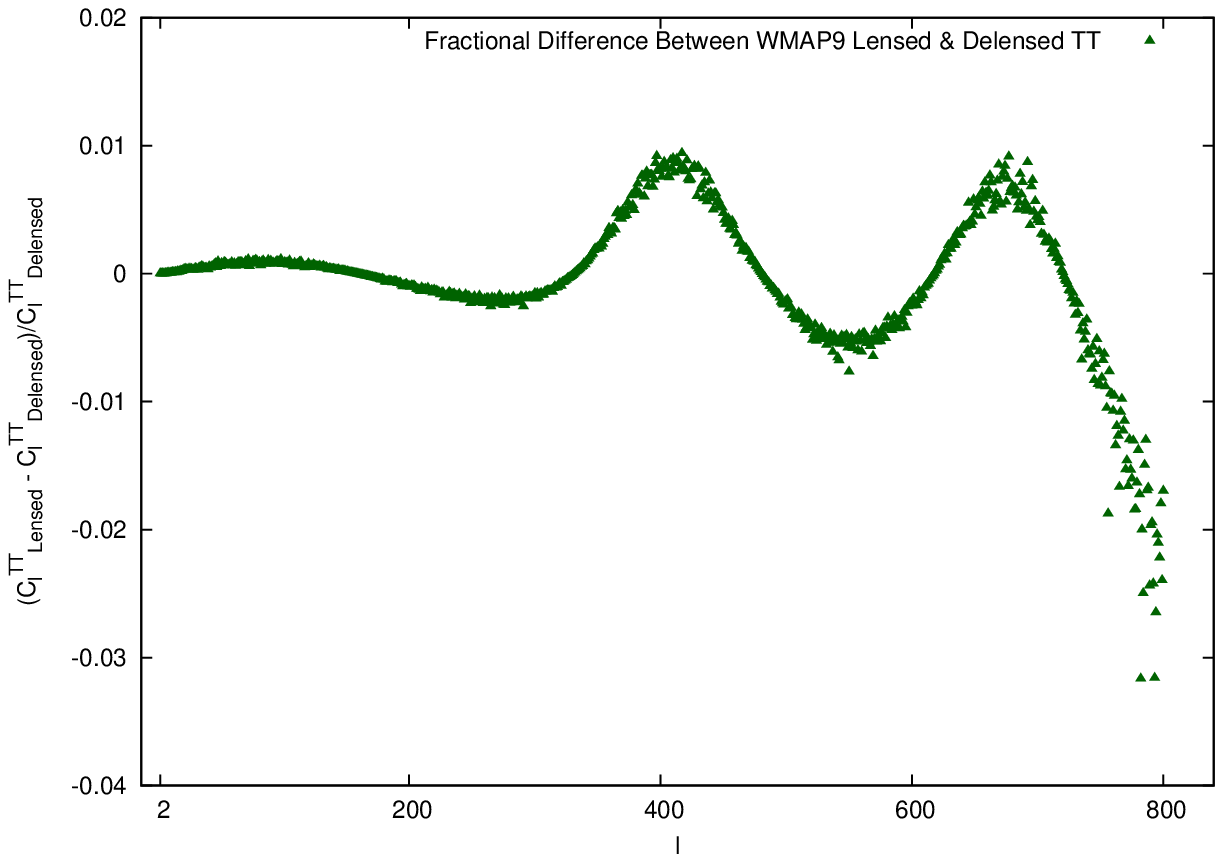}
\end{center}
\caption{The left panel shows the delensed power spectrum as obtained by our above procedure in the full-sky case, using an estimate of the lensing potential power spectrum obtained from CAMB, over plotted on the WMAP 9-year unbinned $TT$ spectra.   
The right panel shows the fractional difference between the unbinned spectra obtained from the WMAP 9-year data and the delensed $TT$ spectra as obtained from our procedure in the full-sky case, this provides an estimate of the lensing contribution to the $TT$ power spectrum.}
\label{fig:wmap9tt}
 \end{figure}

Let us now provide an estimate of the delensed $TT$ spectra, as obtained from our procedure using the WMAP 9-year unbinned $TT$ spectra, and an estimate of the 
lensing potential power spectrum obtained from CAMB using the WMAP 9-year cosmological parameters. The results are shown in Fig.\ref{fig:wmap9tt}. In the left panel, 
the delensed spectra are over plotted on the WMAP 9-year unbinned $TT$ spectra. In the right panel, the fractional difference between the lensed and delensed spectra 
is plotted. It can be seen that the lensing contribution is at the level of $2-3\%$ for values of $\ell \lesssim 800$.

\begin{figure}
\begin{center}
\includegraphics[width=8.5cm, height=8.5cm, angle = -90]{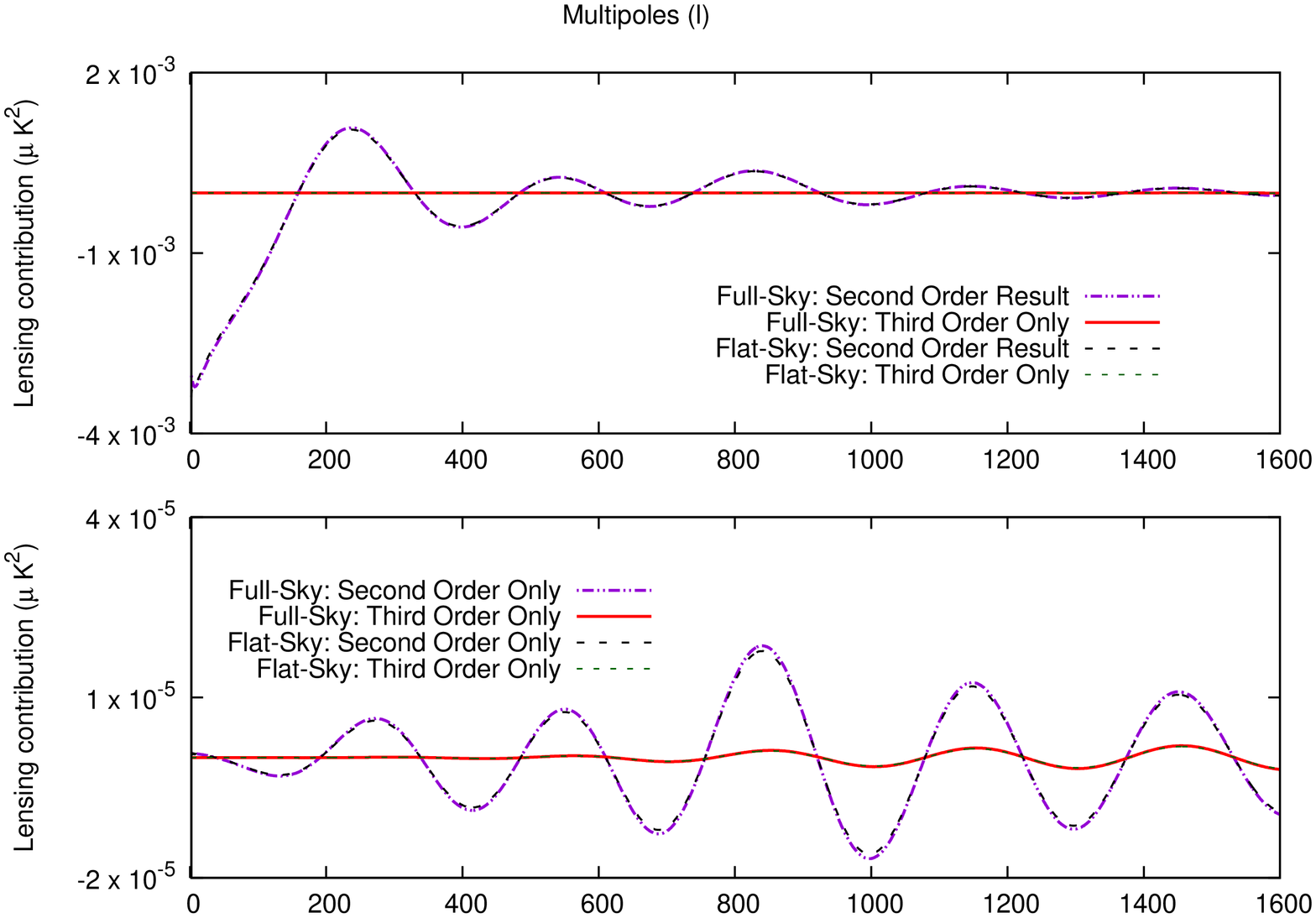}
\includegraphics[width=8.5cm, height=8.5cm, angle = -90]{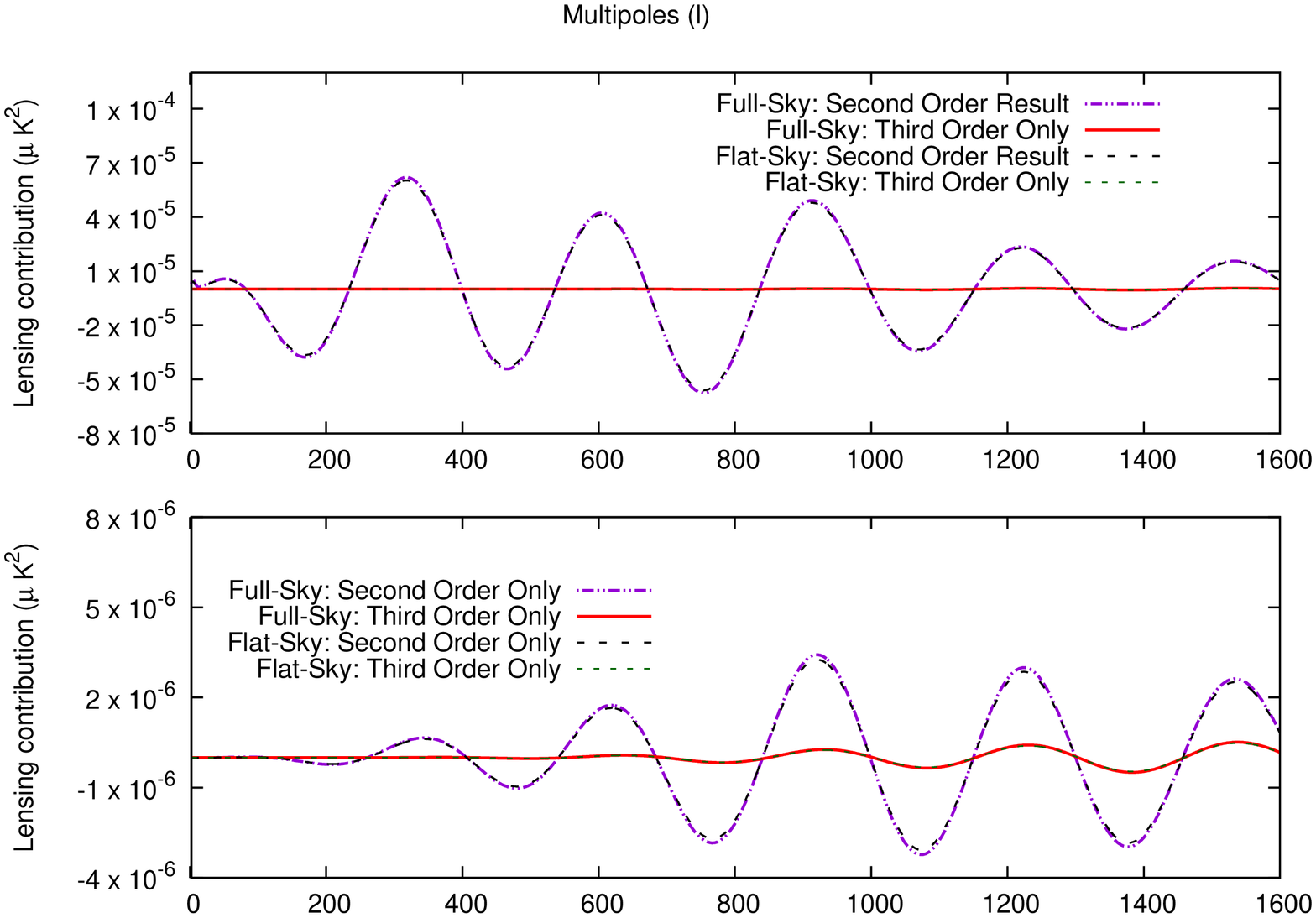}
\end{center}
\caption{The left panel shows the lensing contribution to the CMB $TT$ power spectrum in units of 
$\mu K^2$. In the top half, the lensing contributions from $(\delta K)+(\delta K)^2$ and $(\delta K)^3$ are shown  for both the flat-sky and the full-sky. In the bottom half, only the second- and third-order contributions are plotted, to provide an estimate of the error involved in neglecting the third order term. It is clear that the negligence of the third order kernel matrix does not incorporate 
any significant error in our analysis. The right panel shows the corresponding lensing contributions for the case of the $TE$ spectrum, assuming zero intrinsic $B$-mode power.}
\label{fig:error1}
\end{figure}

\begin{figure}
\begin{center}
\includegraphics[width=8.5cm, height=8.5cm, angle = -90]{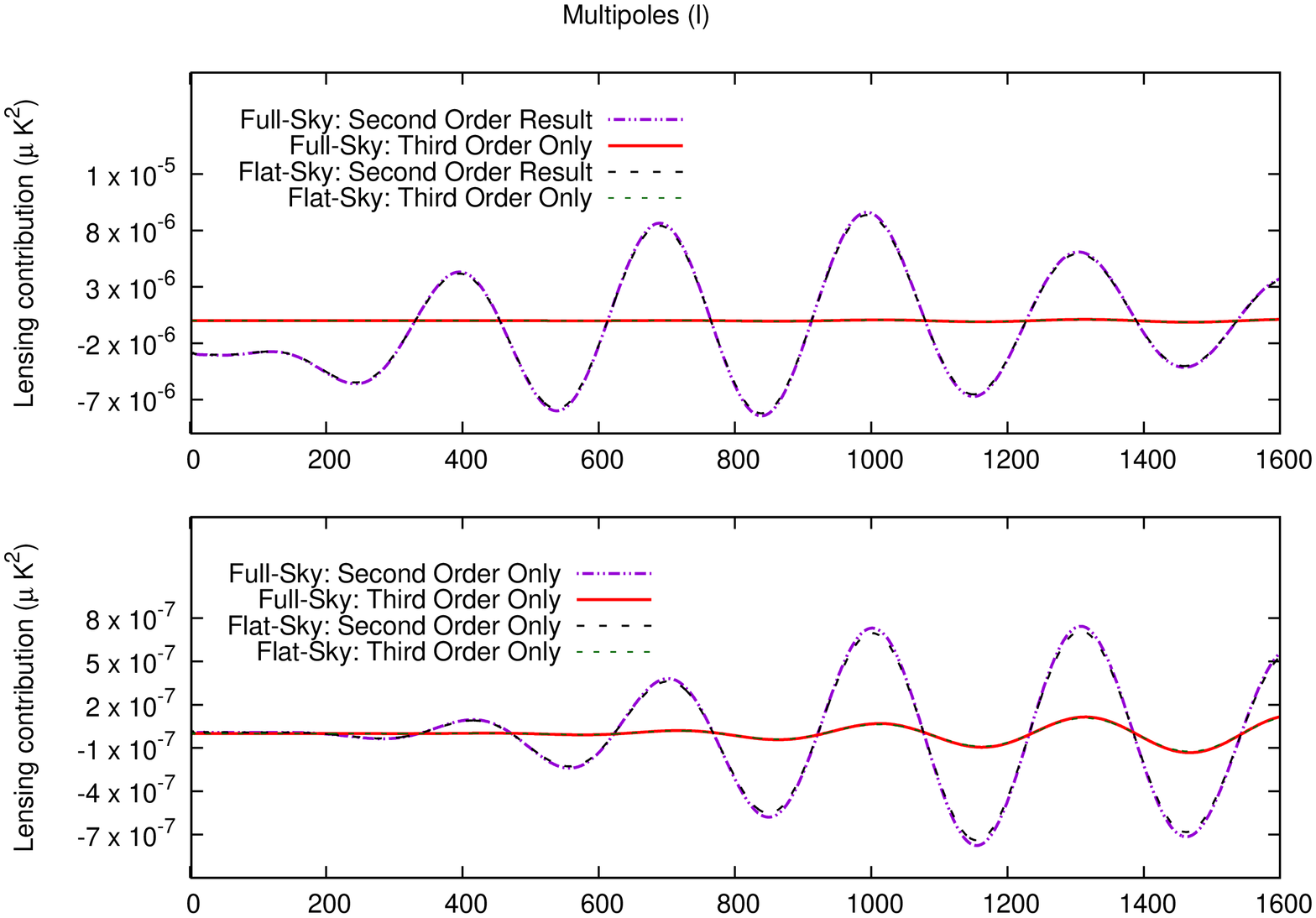}
\includegraphics[width=8.5cm, height=8.5cm, angle = -90]{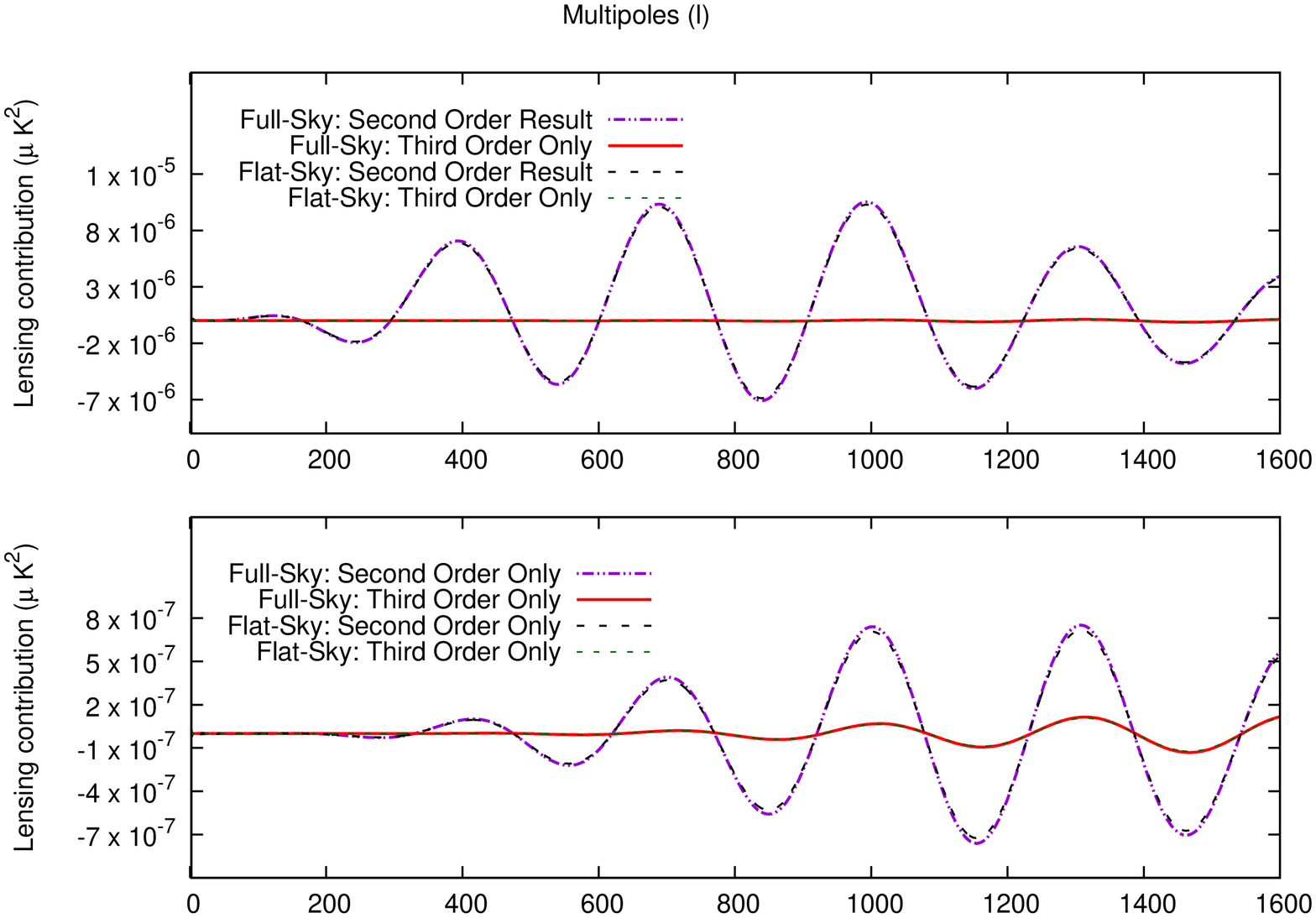}
\end{center}
\caption{The left panel shows the lensing contribution to the $C^{+}$ spectrum in units of 
$\mu K^2$. In the top half, the contributions from $(\delta K)+(\delta K)^2$ and $(\delta K)^3$ are shown 
separately for both the flat-sky and the full-sky. In the bottom half, only the second- and third-order contributions are plotted, to provide an estimate of the error involved in neglecting the third order term. It is clear that the negligence of the third order kernel matrix does not incorporate 
any significant error in our analysis. The right panel shows the corresponding lensing contributions for the case of the $C^{-}$ spectrum.  Here, as in the previous figure, we have assumed zero intrinsic $B$-mode power.}
\label{fig:error2}
\end{figure}

Finally, we furnish a brief estimate of the possible sources of error in our analysis. In the flat sky approximation, as pointed out 
in \cite{Lewis}, 
the error in ignoring terms beyond the second order in $A_2 (\beta) $ is extremely small, of the order of $10^{-4}$. 
The Taylor series expansion for evaluating 
the unlensed spectra, Eqns. \eqref{ultem}, \eqref{ulplus}-\eqref{ulcross} and Eqns.\eqref{sultem}-\eqref{sulcross}, 
ignores terms of the order of $(\delta K)^3$. This introduces an error 
of the order of $(\delta K)^3$, which we find to be at least four orders of magnitude below the contribution from the combination 
of the first and second order terms, for $\ell\lesssim 500$. As we go to higher $\ell$ values, the third order contribution increases, but it still 
remains at most $10^{-3}-10^{-4}$ of the first- and second-order contributions, for about $\ell\lesssim 1200$, as can be seen from Figs.\ref{fig:error1} 
and \ref{fig:error2}. The fractional contribution of the $(\delta K)^3$ term is of the order of $10^{-3}$ for the $TT$ spectrum at large multipoles 
$(\ell\gtrsim 1200)$, but it remains
less than $10^{-4}$ for the other CMB spectra. 
 In the full-sky analysis,
we have neglected the $A_0(\beta)$ terms entirely as they are very small (of the order $<10^{-4}$) \cite{Lewis}. In this case also, the 
contributions from the $(\delta K)^3$ terms are 
tiny, as can be seen from Figs.\ref{fig:error1} and \ref{fig:error2}. Hence, neglecting the $(\delta K)^3$ 
terms does not really incorporate 
significant error in our analysis. Further, our work did not take into account the effect of the non-linear evolution of the lensing 
potential which may also
incorporate some additional error. However, non-linearities are increasingly important only in the very small scale regime. The integrated effect of the above errors leads to the overall accuracy of our analysis, estimated to be 
of the order of $0.33\% - 0.92\%$ at the multipoles around $\ell \sim 1500$-$1600$ by the calibration technique described above. A slightly more accurate result might possibly be obtained by using the full second order expressions for 
the lensed correlation functions in the full 
sky, as provided in Appendix C of Ref. \cite{Lewis}, but would increase the computational expense. 

\section{Conclusions}\label{dis}
In this article, we have provided a new method to extract the intrinsic CMB power spectra from the lensed ones by 
applying a matrix inversion technique. After a through formal development of the theoretical framework, we use a FORTRAN90 code to compare our results with 
CAMB. We find that for $\ell\lesssim 1000$, both the results are almost identical with the fractional deviation being of the order of $10^{-4}$ or so, but as we go beyond that, 
the difference becomes about $0.33\%$ - $0.92\%$ around $\ell\sim 1600$. Thus, this new technique can very well serve as a first step towards direct 
reconstruction of intrinsic CMB power spectra from the lensed map.

To apply our methodology, the knowledge of lensing potential is necessary. The lensing potential can be reconstructed once we have the full CMB polarization 
data, \textit{i.e.}, both $E-$ and $B-$ mode. Though we do not address the lensing potential reconstruction, it may be possible to 
construct the CMB lensing potential from the complete polarization data alone.   
There is a confusion between the CMB $E$- and $B$- polarization modes due to their mixing in the presence of weak lensing. 
Nevertheless, using our matrix inversion technique, it may be possible to obtain a 
handle on the intrinsic $B$-mode power by subtracting the lensing effect due to
 $E$-modes, once the lensing potential is known. Of course, as we have 
 clearly mentioned, this work is just the first step towards this reconstruction. Here, obtaining the unlensed CMB spectrum and the demonstration of 
deconvolution of the lensing effect has been done in the ideal situation, without taking into account noise in the measured $C_{l}$'s, 
uncertainties in the transfer function and in the primordial power spectrum used in the transfer function as 
well as errors coming from other sources as discussed in Section \ref{num}. In the realistic situation, the above uncertainties need to 
be incorporated. We hope to address the reconstruction of the lensing potential from the lensed CMB polarization data (with an estimate for large-scale $B$-mode power generated using CAMB) in a future work.
A possible further direction in which this work may be taken is in the increase of 
accuracy in determining various 
cosmological parameters, which we hope to address in future.

\section*{Acknowledgements}
We thank Tarun Souradeep for discussions and useful suggestions.
BKP thanks the Council of Scientific and Industrial Research (CSIR),
India for financial support through Senior Research Fellowship
(Grant No. 09/093 (0119)/2009). The research of HP is supported by the SPM research grant of CSIR, India.
SP thanks ISI Kolkata for computational support through a research grant. We also acknowledge the use of publicly available code CAMB (http://www.camb.info) 
for calibration of our analysis.

\label{lastpage}

\end{document}